\documentclass[preprint,showpacs,preprintnumbers,amsmath,amssymb]{revtex4}

\usepackage{graphicx}
\usepackage{dcolumn}
\usepackage{bm}
\usepackage{color}
\def\be{\begin{eqnarray}}\def\ba{\begin{eqnarray}}
\def\ee{\end{endqnarray}}\def\ea{\end{eqnarray}}
\def\no{\nonumber\\}

\begin{document}


\title{Density driven symmetry breaking in holographic superconductors}

\author{Youngman Kim}
\email{ykim@apctp.org}
 \affiliation{Asia Pacific Center for Theoretical Physics\\ and Department of Physics, Pohang University of Science and Technology,
Pohang, Gyeongbuk 790-784, Korea}

\author{Yumi Ko}%
 \email{koyumi@sogang.ac.kr}
\affiliation{%
Center for Quantum Spacetime, Sogang University, Seoul 121-742, Korea
}%

\author{Sang-Jin Sin}
 \email{sjsin@hanyang.ac.k}
\affiliation{
Department of Physics, Hanyang University, Seoul 133-791, Korea
}%

\date{\today}

\begin{abstract}
We study the density driven symmetry breaking in holographic superconductors by considering the positive mass squared case. We show that even for the positive $m^2$, a scalar condensation still forms, provided the chemical potential is high enough.
As  $m^2$ increases, the phase space folds due to the non-linearity of the equations of motion, and  two nearby points in the phase space can represent  symmetry breaking  and preserving configurations respectively.
 The phase space defined by the set of initial conditions of field variables at the horizon undergoes a
 non-linear radial evolution to result in the phase space folding, a characteristic phenomenon in a non-linear system.
We then calculate the specific heat, which characterizes superconductors and has been measured in experiments.
We observe a discontinuity in the specific heat at the transition point  and compare our results with experimentally observed numbers.
The electrical  conductivity  for various $m^2$ is also calculated.
\end{abstract}

\maketitle

\section{Introduction}

It has been suggested that some of phenomena of strongly  interacting systems
  can be described by the  AdS/CFT correspondence \cite{Maldacena:1997re}.
While the relevance of the gauge/string duality  to the QCD has been  much discussed in the context of heavy ion collisions \cite{son, SZ,Mateos:2007ay} and hadron physics, its relevance to the electron system is rather new and still an open  issue.  However, testing the applicability with a model in the bulk and the phenomena at the boundary is a very exciting possibility.
In this context, it has been attempted to use the AdS/CFT correspondence to describe certain condensed matter systems such as the quantum Hall effect \cite{Hartnoll:2007ai}, Nernst effect \cite{Hartnoll:2007ih, Hartnoll:2007ip, Hartnoll:2008hs},  superconductivity \cite{Hartnoll:2008vx, Ammon:2009fe, Horowitz:2008bn}, and  fractional quantum Hall effect \cite{Fujita:2009kw}.

In Ref. \cite{Hartnoll:2008vx}, a model of a strongly coupled system which develops superconductivity was suggested  based on the holography, which is an Abelian Higgs model in  3+1 dimensional anti-de Sitter(AdS) spacetime. In this model a complex scalar field, as a   dual of the ``Cooper pair'' operator at the boundary, is introduced whose vacuum expectation value is the order parameter of the superconductivity. The symmetry breaking was induced  by considering the negative mass squared of the scalar field. In AdS spacetime, the negative mass squared is allowed due to the Breitenlohner-Freedman(BF) bound. In this case, the dual operator coupled to the scalar field  at the boundary has a weight $\Delta=1$  or $\Delta=2$. Recall that for a scalar field in AdS$_{d+1}$ spacetime, $(mL)^2=\Delta(\Delta-d)$. The cases with $\Delta=3/2$ and 3 were considered in Ref. \cite{Horowitz:2008bn}. The hydrodynamic nature was studied in a theory with large conformal operators, $1/2\le\Delta \le 5$, in Ref. \cite{Yarom}.

In 2+1 dimensions, a fermion bilinear (Cooper pair) has naive dimension 2, therefore its dual scalar field should have a tachyonic mass, leading to the instability for the symmetry breaking. However,  the anomalous dimension is usually non-zero and can be large, so we are forced to ask what happens if the mass squared  is positive. Another motivation is the neutron superfluidity in the neutron star.
Since the neutron consists of three quarks, its creation operator is a composite operator with high conformal dimensions and so is the Cooper pair of the neutrons, naively $\Delta=9$ for the pair.  Again, one is led to the question of the positive mass squared case.

When the scalar field has a positive mass squared, one may expect that the mechanism of the symmetry breaking is also removed. However,  in the presence of the matter, there is a density driven instability leading to the symmetry breaking as the effective mass of the complex boson in flat spacetime is given by $m_{eff}^2=m^2-\mu^2$. Here $\mu$ is the chemical potential.

In warped spacetime,  we  can still expect  the same mechanism  at work. This is already pointed out in Ref. \cite{Gubser:2008wv}  from the fact that the fluctuation of the scalar field has an effective mass 
\begin{equation}\label{meff}
m_{eff}^2=m^2- |g^{tt}|A_t^2,
\end{equation}
where  $A_t$ is the electrostatic potential whose value at the boundary is the chemical potential.
 Notice that the regularity requires  $A_t=0$ at the horizon. As a consequence, the second term of Eq. (\ref{meff}) goes to zero at the horizon  in spite of the divergence of $g^{tt}$.  On the other hand, near the boundary, $g^{tt}$ is zero hence it eliminates the effect of the chemical potential. Therefore, the density effect should be less effective than the flat space case, and it is not clear whether the density driven symmetry breaking mechanism is still working. The present work is to address this question in detail.

 In fact, our results show that for a small positive mass squared, the result is not much different from the negative case studied in Refs. \cite{Hartnoll:2008vx, Basu:2008st, Herzog:2008he, Hartnoll:2008kx, Horowitz:2008bn, Gubser:2008wv}.
However,  as the mass squared increases, the phase space folds so that two very close boundary conditions  can give completely opposite states: one  symmetry  broken state, and another  symmetric state.

The specific heat is one of the experimental observables for superconductors, and it has two interesting limits.
At a temperature far below the transition temperature $T_c$, at which the superconductivity disappears,
the specific heat is exponentially small due to the pairing gap, while  it exhibits a discontinuity very near $T_c$.
We calculate the specific heat with various $m^2$ to see if large conformal dimensions change the properties, focusing on the normalized magnitude of the discontinuity. Then we compare our results with experimentally observed numbers of some real materials.
Finally we calculate the electrical conductivity for $m^2>0$ and $m^2<0$.

\section{Abelian Higgs Model and Holographic Superconductors}%
We use the Abelian Higgs model in AdS$_{d+1}$ Schwarzschild black hole background to study the holographic superconductors. The action of the model is given by
\begin{equation}
S = -\int  d^{d+1} x \sqrt{-g} \left( \frac{1}{4} F^{ab} F_{ab} +  |\partial \Psi - i A \Psi|^2 +m^2 |\Psi|^2\right)\, ,\label{action}
\end{equation}
where the bulk Maxwell field strength is $F_{ab}=\partial_a A_b -\partial_b A_a$, and $\Psi$ is a charged complex scalar field.
We work with the following (d+1)-dimensional background metric:
\ba
ds^2 = \frac{1}{z^2} \left( - f(z) dt^2 + \frac{dz^2}{f(z)} + d {\vec x }^2\right) ,\quad
f(z) = 1 - \left( \frac{z}{z_0}\right)^d ,
\ea
where the horizon is at $z=z_0$.  The Hawking temperature of the black hole is given as
\begin{equation}
T = \frac{d }{4 \pi z_0}.
\end{equation}
Note that we will work in the probe limit: the Maxwell field and the scalar field do not give any back-reaction to the background metric.
Here we take the form of the bulk scalar field as $\Psi = \frac{1}{\sqrt{2}} \psi e^{i \phi}$ and consider the gauge transformation $A_a \rightarrow A_a + \partial_a\phi$.
Then the equations of motion for static and translational invariant  fields of interest read
\begin{eqnarray}
\psi '' + \left( \frac{f'}{f} - \frac{d-1}{z}\right) \psi' +\left( \frac{A_t^2}{f^2} - \frac{m^2}{z^2 f}\right) \psi = 0\, , \no
A_t'' + (3-d) \frac{A_t'}{z} - \frac{\psi^2 }{z^2 f}\, A_t= 0\, . \label{EoMz}
\end{eqnarray}
To solve the equations of motion, we specify appropriate boundary conditions as follows: at the horizon,
\begin{eqnarray}
 \quad A_t (z_0) = 0 ,\quad \psi( z_0 ) = -\frac{d \,z_0}{m^2} \psi'(z_0), \label{BCh}
\end{eqnarray}
 and at the boundary, i.e. $z\rightarrow 0$,
\begin{eqnarray}
\psi(z) \!\!&=&\!\! \Psi_1 z^{d-\Delta} + \Psi_2 z^{\Delta} + \cdots \nonumber\, ,\\
A_t (z) \!\!&=&\!\! \mu - \rho z^{d-2}+\cdots.\label{BCb}
\end{eqnarray}
We will fix the chemical potential $\mu$ and thus work in the grand canonical ensemble.  For the boundary value of the scalar field $\psi(z)$,  we will choose $\Psi_1=0$.
When both $\Psi_1$ and $\Psi_2$ are normalizable modes we need to make a choice of either $\Psi_1 = 0 $ or $\Psi_2 = 0$ \cite{Hartnoll:2008vx}.
As we increase $m^2$,    $\Delta$ increases and $\Psi_1$ becomes non-normalizable mode,  which corresponds to a source term. We choose $\Psi_1=0$ to impose the source free condition  which is necessary to study the spontaneous symmetry breaking rather than induced breaking.

Notice  that the equations of motion Eq. (\ref{EoMz}) are invariant under the   re-scaling:
\ba
z \!\!&=&\!\! z_0 {\bar z },\quad  A_t = \frac{{\bar A}_t}{z_0},
\ea
with $\psi(z)$ fixed. At the boundary, we then have
\ba
&&{\bar A}_t ({\bar z}) \sim {\bar \mu} - {\bar \rho}\, {\bar z}^{d-2}\, ,\no
&&\psi({\bar z}) \sim{\bar \Psi}_1 \,{\bar z}^{d-\Delta} + {\bar \Psi}_2 \,{\bar z}^{\Delta}\, ,
\ea
where
\ba
\bar\mu=\mu z_0,\quad \bar\rho=\rho z_0^{d-1},\quad \bar \Psi_1 =\Psi_1 z_0^{d-\Delta},\quad{\rm and}\quad
\bar\Psi_2=\Psi_2z_0^\Delta\, .
\ea

There have been many studies on the properties of the holographic superconductors in 2+1 and 3+1 dimensions \cite{Hartnoll:2008vx, Basu:2008st, Herzog:2008he, Hartnoll:2008kx, Horowitz:2008bn, Gubser:2008wv}.  We note that in the literatures, $m^2$ has been assumed to be negative to guarantee the symmetry breaking and the scalar field condensate. In Ref. \cite{Horowitz:2008bn}, the massless case was also considered.

\section{The Phase Space Folding }
In this section, we extend the previous studies on the holographic superconductors by considering
 the bulk scalar field with the positive mass, $m^2>0$.
As in the   $m^2\le 0$ cases, we  need to solve the coupled equations of motion in Eq. (\ref{EoMz}) with the boundary conditions, (\ref{BCh}) and (\ref{BCb}) with the choice of $\Psi_1 = 0$ and fixed $\mu$.

 First we define the `phase space' as a set of possible values of pair $(\psi' ,\, A'_t  )$ at the horizon $z_0$,  which evolves radially to the boundary  $z=0$. To understand this properly, it is useful  to consider some details of the numerical integration.  If we choose to integrate starting from the horizon to the boundary, we need the values of the fields and their first derivatives at the horizon. However,
half of the actual boundary conditions are imposed at the boundary, $z=0$.
Therefore  we artificially assign arbitrary values of the first derivative of the fields $(\psi'(z_0),\, A'_t(z_0) )$ at the horizon.
Some of them will give the correct boundary conditions given in Eq.(\ref{BCb}) with $\Psi_1=0$ at the boundary.
Such a set of  `initial conditions' giving correct boundary conditions  draws  contours in the plane of $(\psi'(z_0), \,A'_t(z_0) )$.
We define such contours as our  phase space. We use the word ``phase space" to have the same meaning as the configuration space.

\begin{figure}
\centering
\includegraphics[width=7cm]{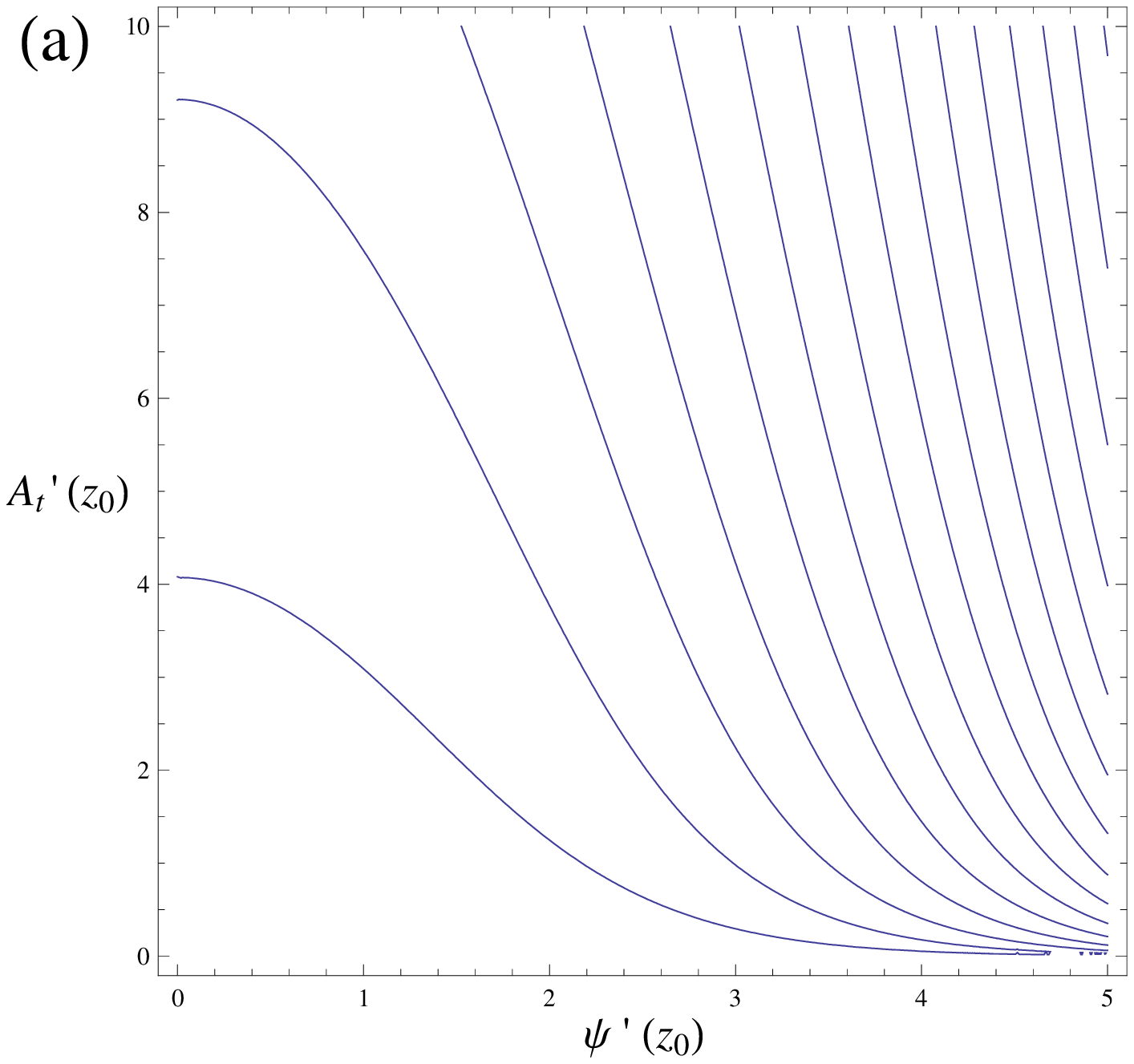}~
\includegraphics[width=7 cm]{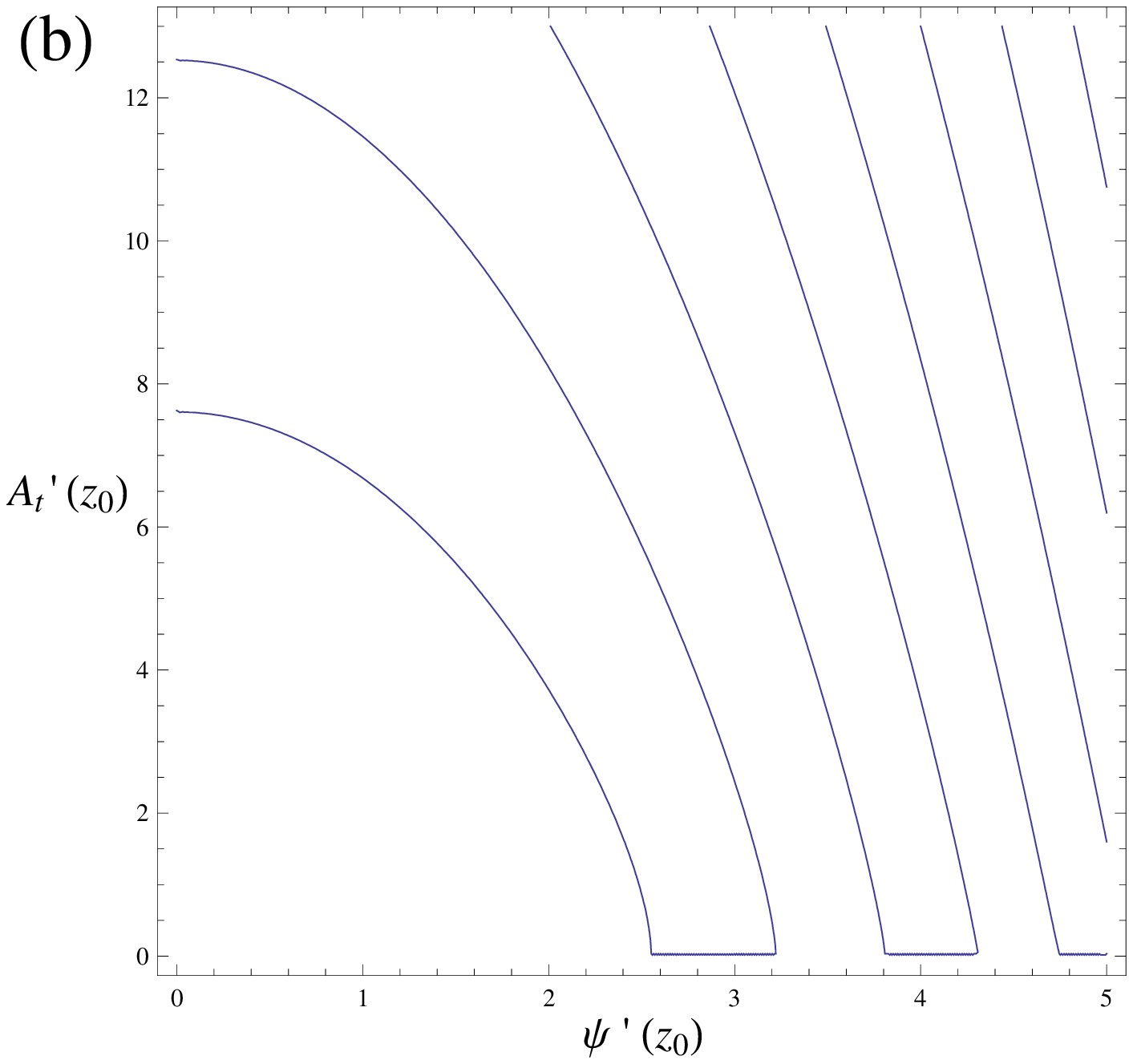}~
\caption{ Phase space as the contours of $(\,\psi'(z_0), A_t'(z_0)\,)$  satisfying $\Psi_1=0$. (a)  $m^2=-2$, (b) $m^2=0$.
Each contour line satisfies the boundary condition $\Psi_1 = 0$.  }\label{contourz}
\end{figure}

\begin{figure}
\centering
\includegraphics[width=7cm]{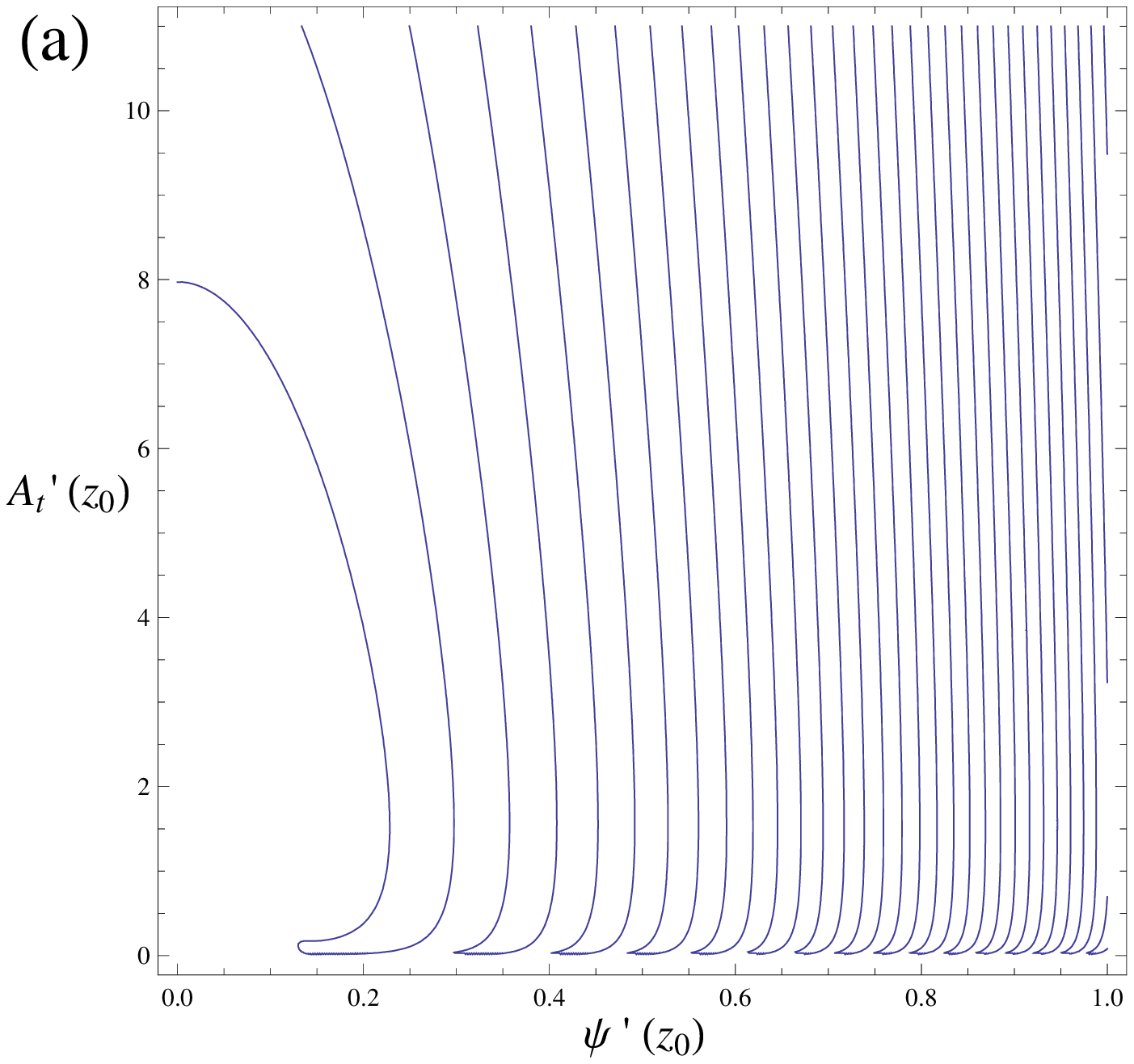}~
\includegraphics[width=7 cm]{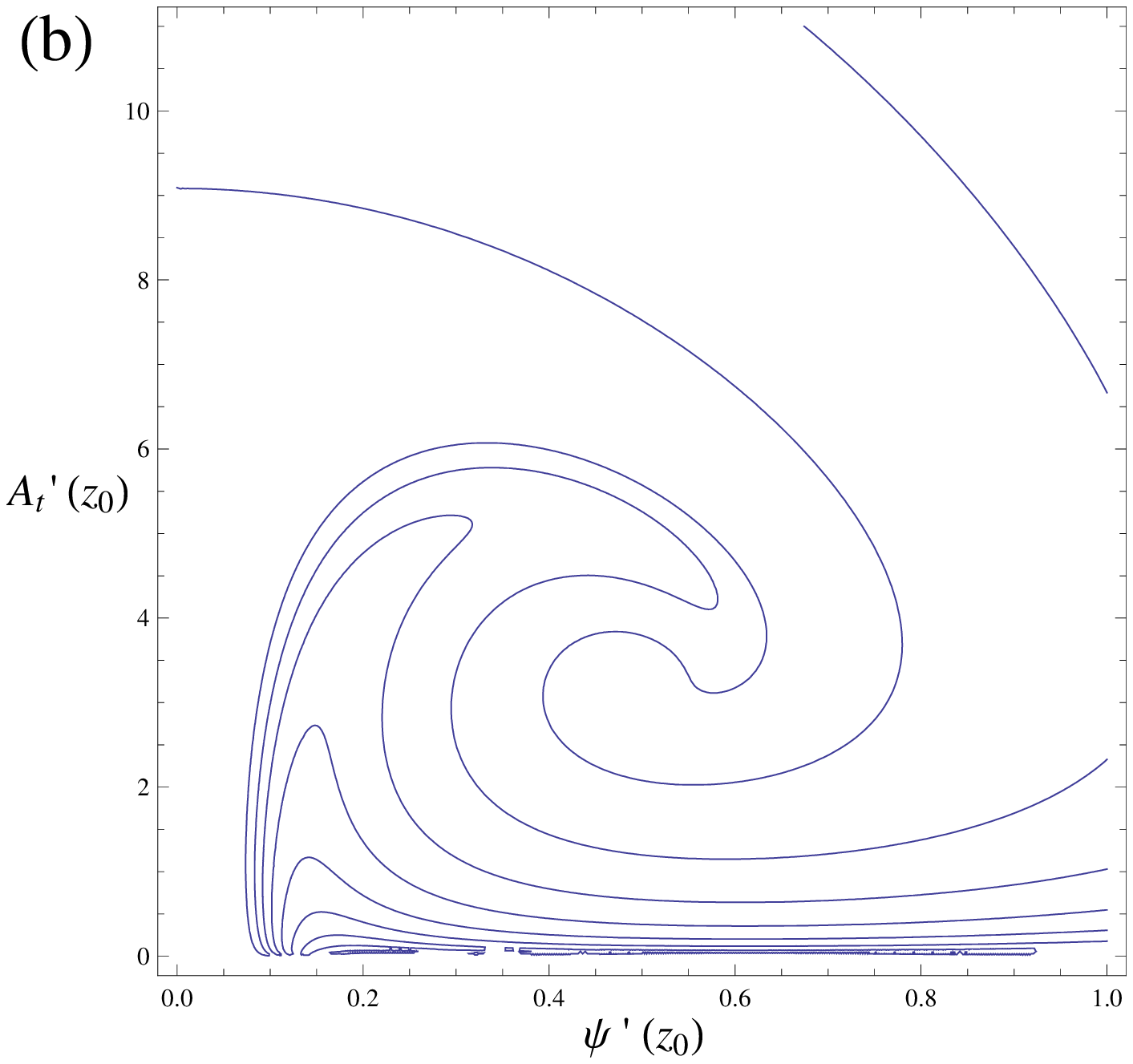}~\\
\includegraphics[width=7cm]{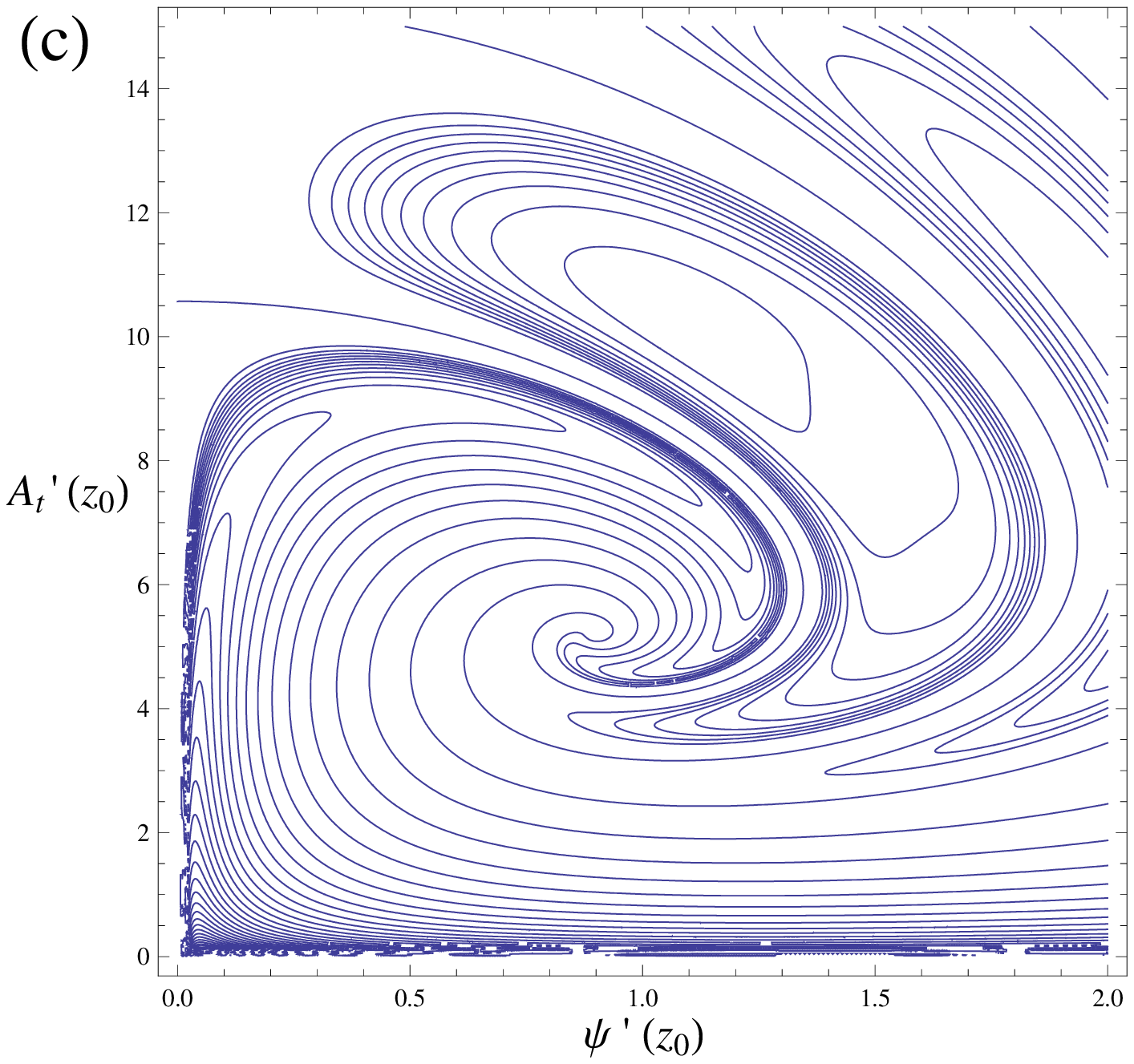}~
\includegraphics[width=7cm]{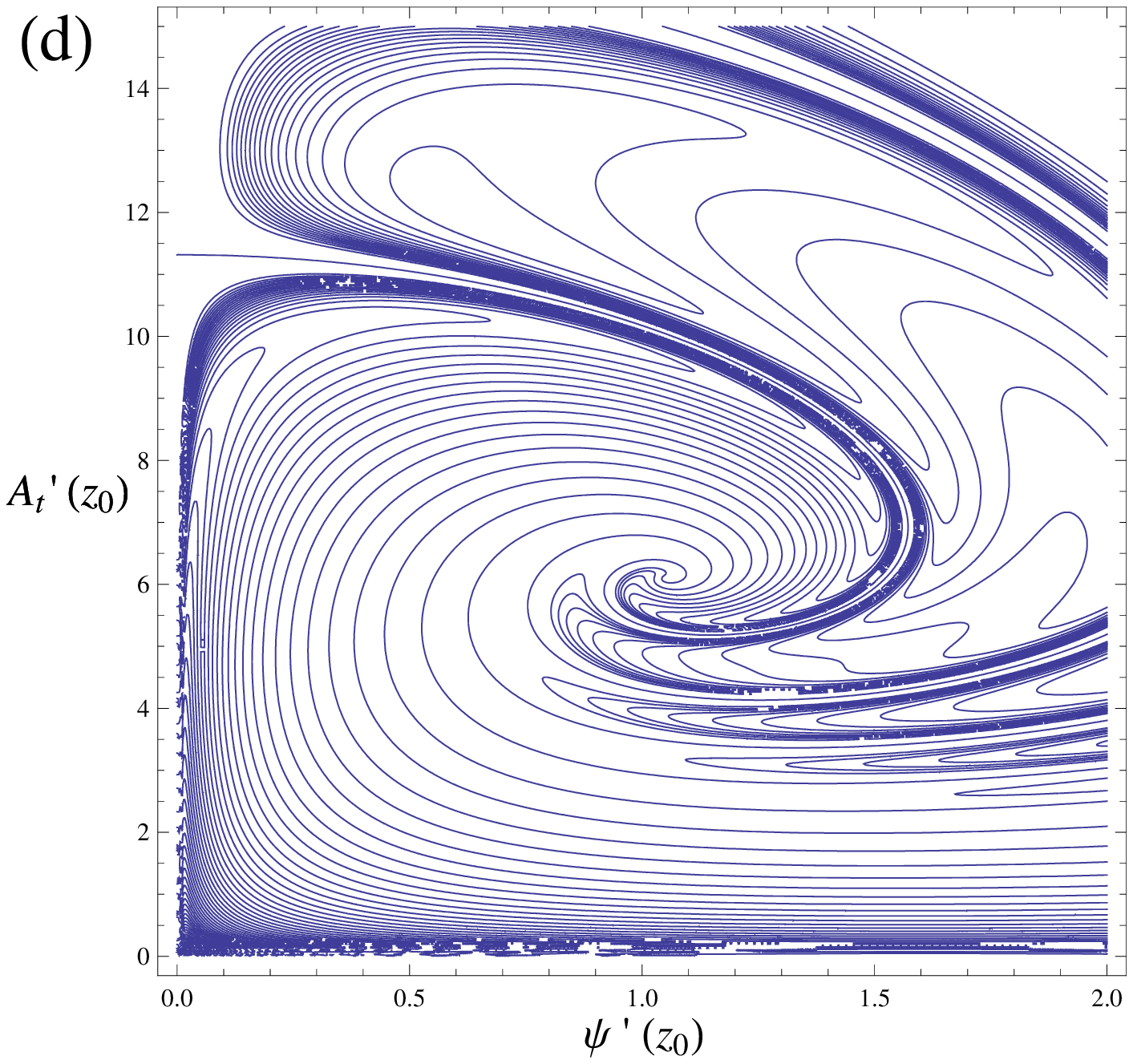}
\caption{ Contours of $(\,\psi'(z_0), A_t'(z_0)\,)$ for various positive $m^2$.
Each contour line satisfies the boundary condition $\Psi_1 = 0$.
(a) $m^2 = 0.31$,
(b) $m^2 = 1.36$,
(c) $m^2 = 3.04$,
(d) $m^2 = 4$,}\label{contourz2}
\end{figure}

For $m^2\le 0$, the phase space is drawn in Fig. \ref{contourz} (a). In Figs. \ref{contourz} and \ref{contourz2}, the contours are for different nodes.  As $A_t'(z_0)$ approaches zero, some contour lines misleadingly look connected to its neighbor but none of them actually meet each other.  

\begin{figure}
\centering
\includegraphics[width=7cm]{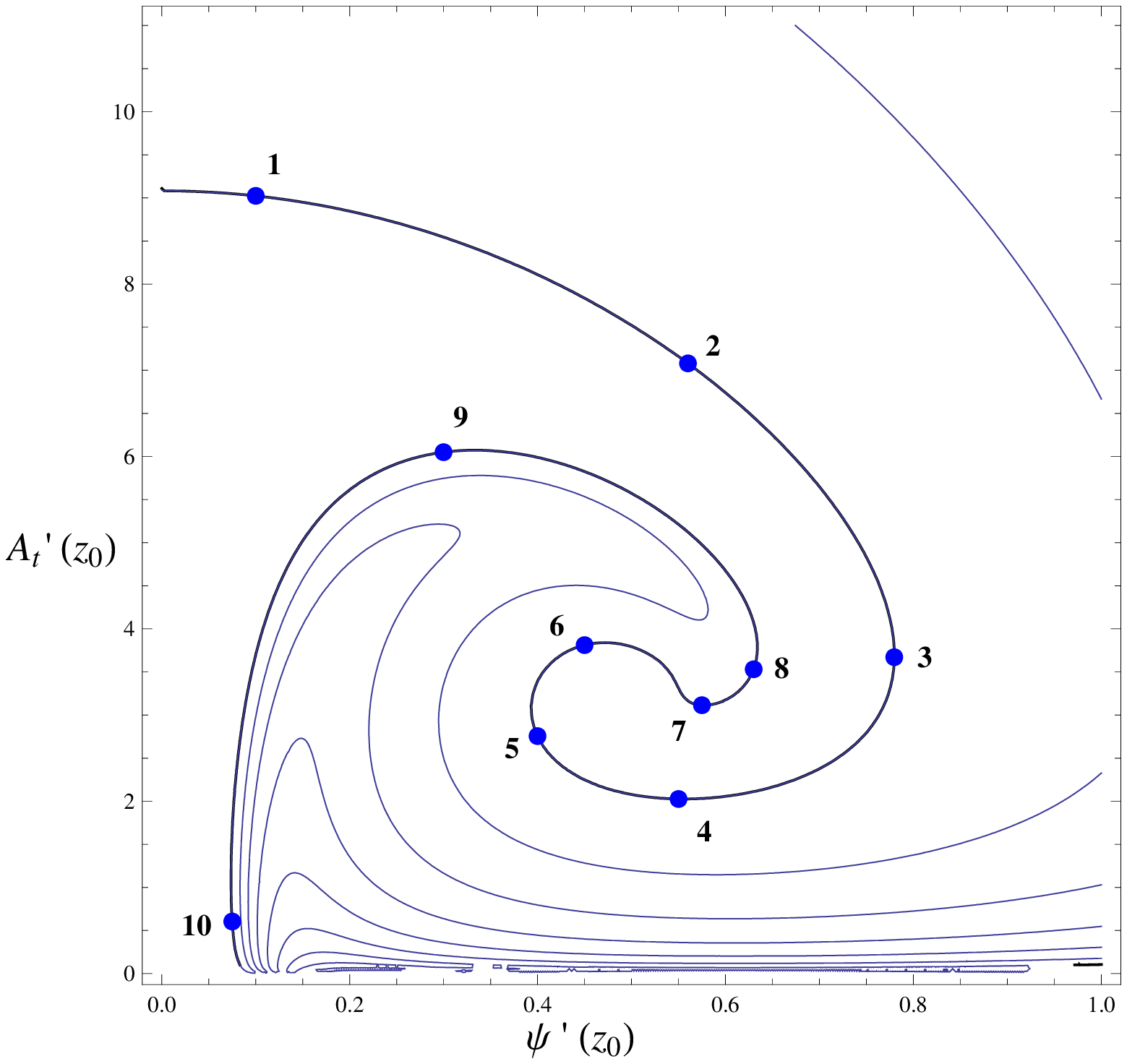}~
\includegraphics[width=9cm]{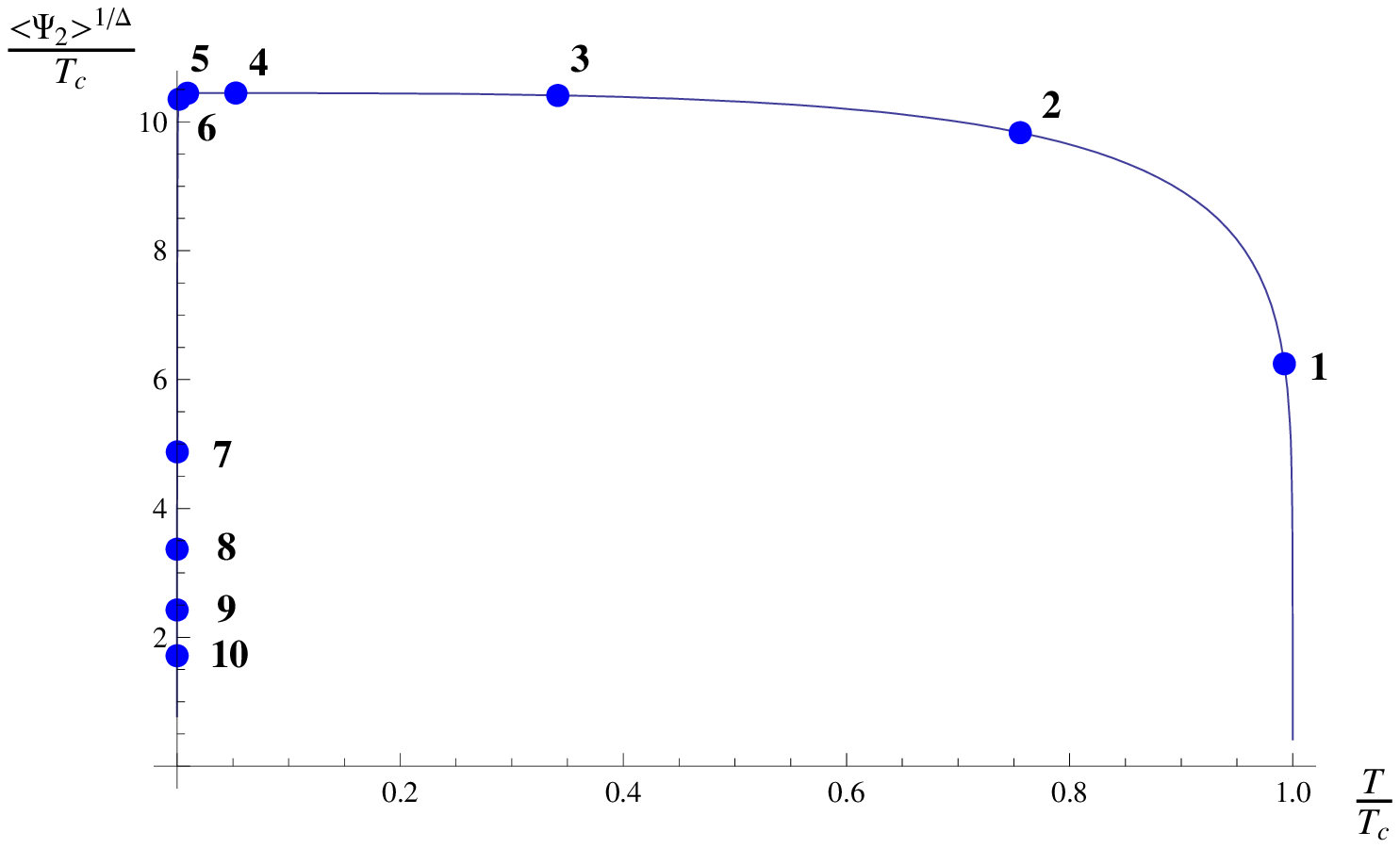}
\caption{The wrapped phase space for $m^2 = 1.36$ and
the condensation. The numbers, 1-10,  were to indicate the correspondence between the configuration and the condensation state. Since $T_c$ is fixed by the chemical potential, the marking numbers  can also be considered as a measure of (increasing) chemical potential for fixed temperature.
}\label{f136}
\end{figure}

\begin{figure}
\centering
\includegraphics[width=9cm]{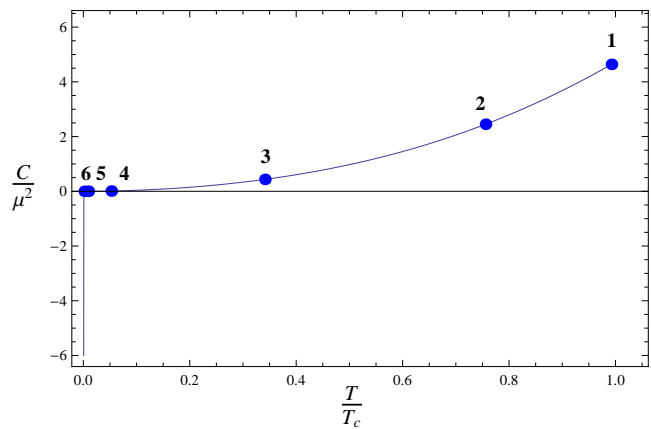}~
\caption{Specific heat for each point on the phase space in Fig. \ref{f136}. The numerical values of the specific heat for the points 7, 8, 9 and 10 are  -1389, -6106, -5581 and -4731 respectively.  }\label{cp136}
\end{figure}

\begin{figure}
\centering
\includegraphics[width= 6cm]{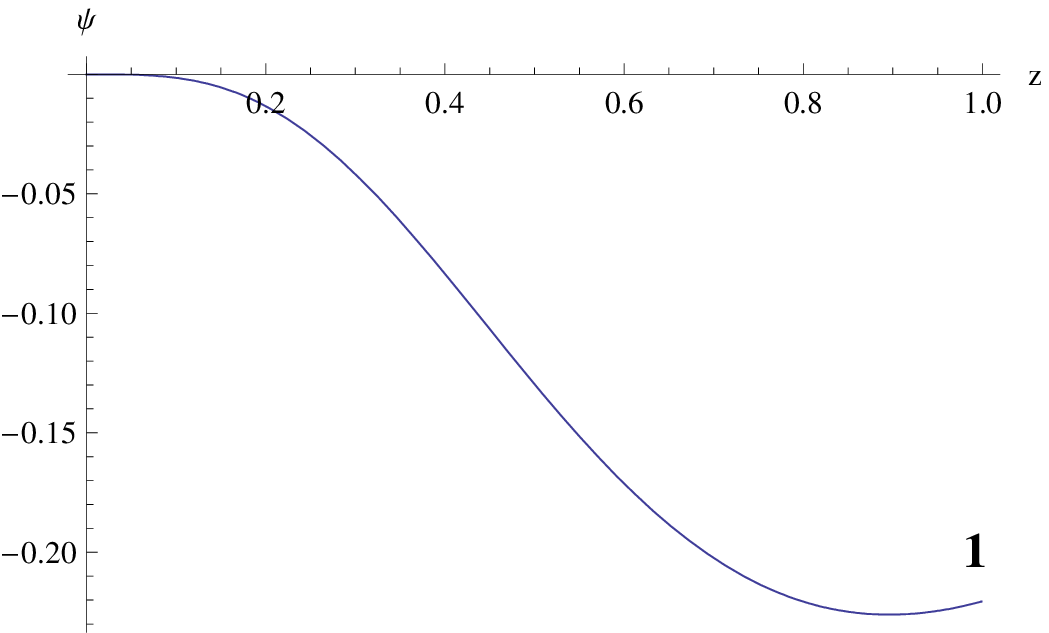}~~
\includegraphics[width=6cm]{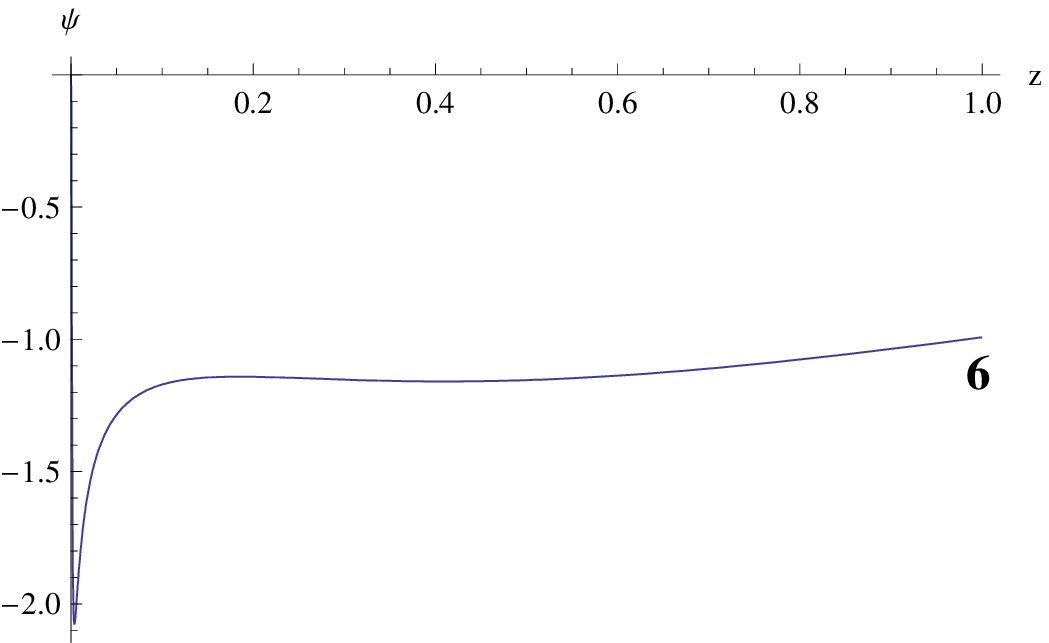}\\
\includegraphics[width= 6cm]{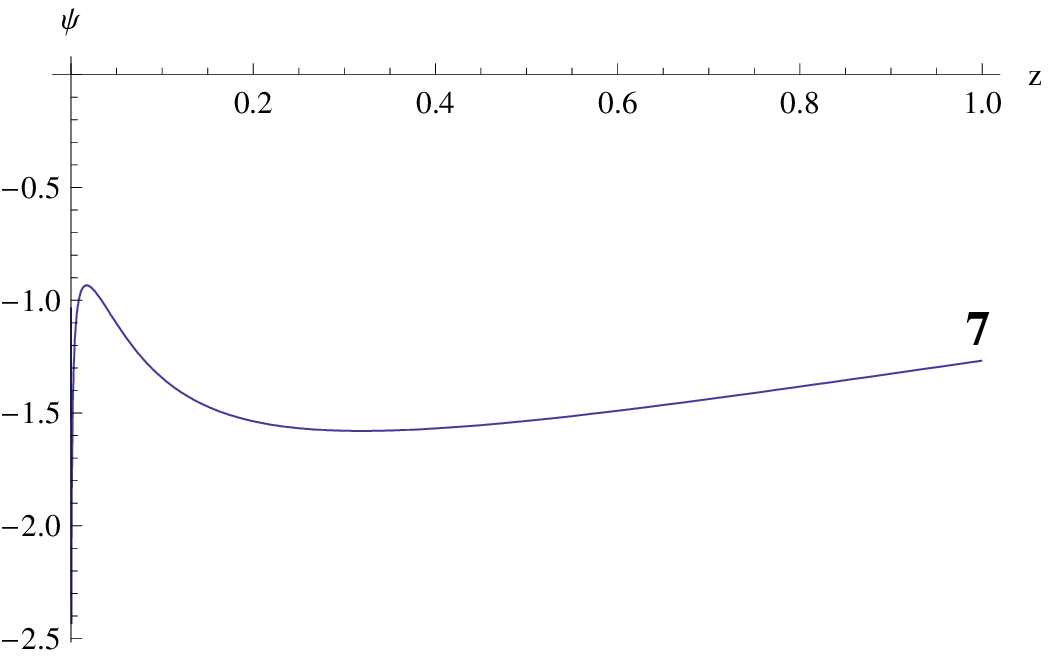}~~
\includegraphics[width=6cm]{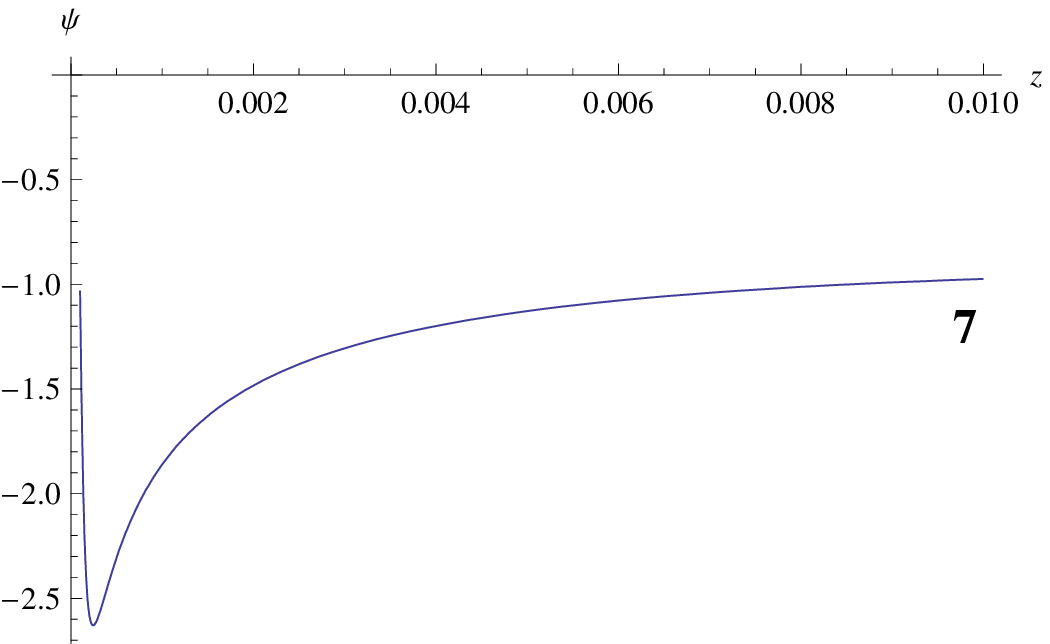}
\caption{Shape of $\psi$ for  a few points of phase space in Fig. \ref{f136}.
The lower right figure is the closer look near $z=0$ of the needle shape in the lower left figure.}\label{w136}
\end{figure}

As the value of $m^2$ increases, the phase space begins to wrap, as seen in  Fig. \ref{contourz2}. In order to see what is happening, we choose a phase space for $m^2=1.36$  and take several points there as seen in Fig. \ref{f136}. We number the phase space points such that  $\mu/T$  increases as the number increases.  Then we calculate the scalar field condensate and the specific heat and draw the shape of the scalar field configuration. See Figs. \ref{f136}, \ref{cp136} and  \ref{w136}  respectively. 

We see that the wrapped part which contains  the points 6-10 is unstable since the specific heat is negative there.
To support this claim further, we also study the shape of the scalar field configuration and draw the result in Fig. \ref{w136}.
For a point in the folded region of the phase space, the corresponding wave function develops a needle like shape which costs
too much energy, making the configuration unrealizable.


 On the other hand, the configurations in the unwrapped part (containing points labeled by 1,2,3,4,5) are regular and realizable. As we increase $m^2$ further, however,  the phase space progresses to fold itself until it  gives an interesting behavior: two infinitesimally close points in the phase space will give completely different characters of the states in the boundary, one ordered and   the other disordered.  At the onset of such an effect, the superconducting and the normal states might  coexist in an arbitrarily mixed fashion. It would be interesting if such a mixing is in chaotic fashion.  

\begin{figure}
\centering
\includegraphics[width=8cm]{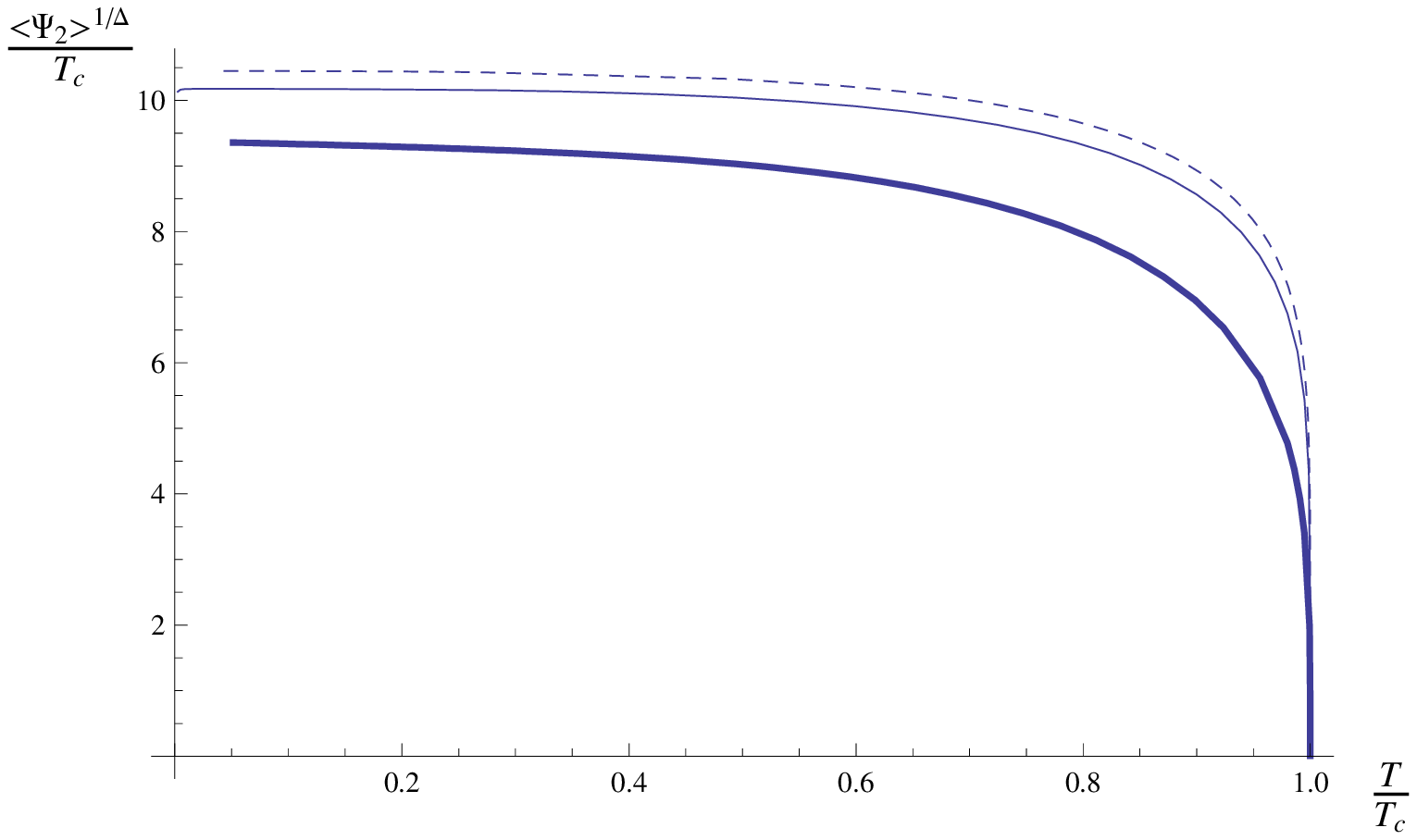}~
\includegraphics[width=8 cm]{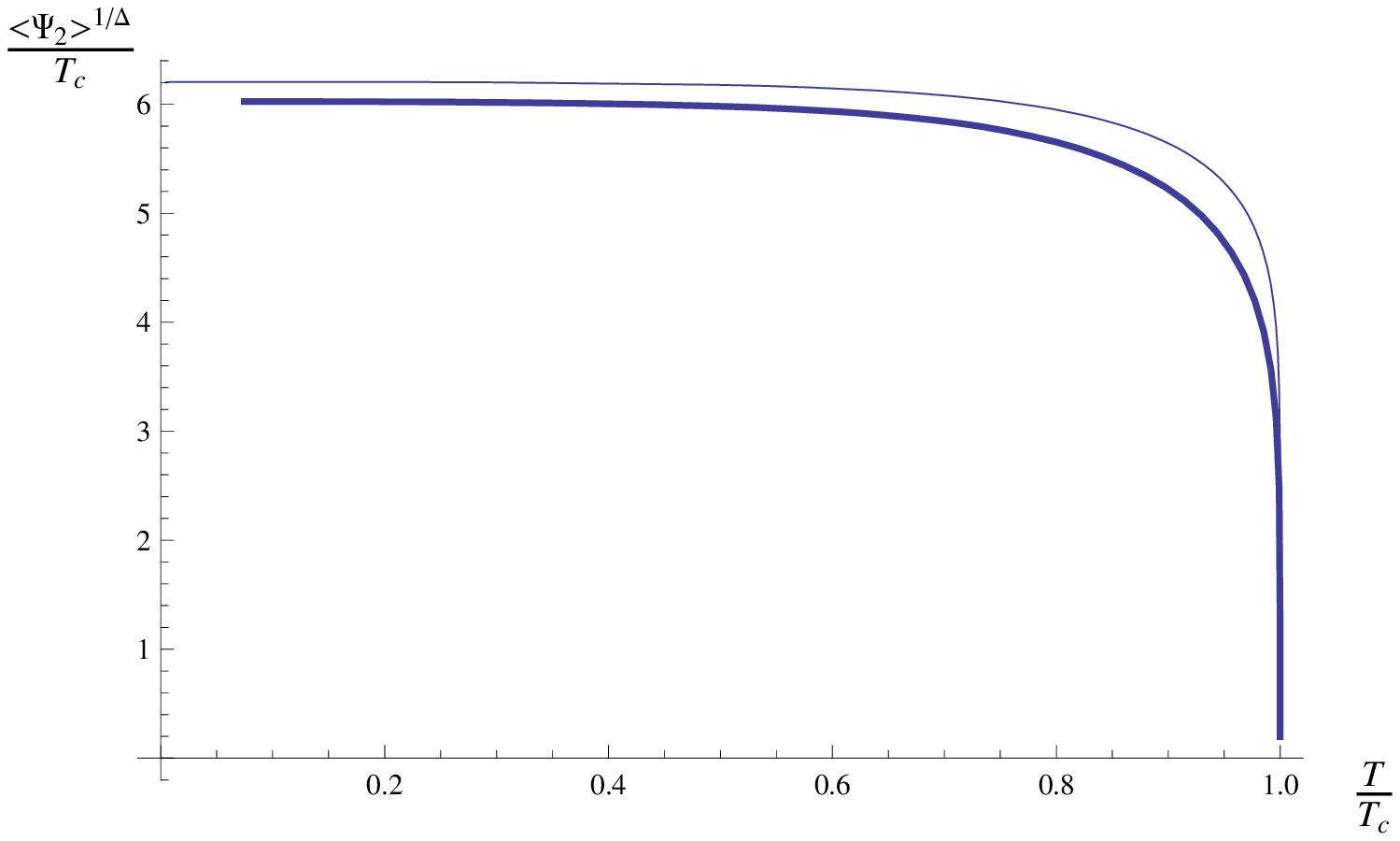}~
\caption{The condensation $\Psi_2$ in d=2+1 (left) and d=3+1 (right).
Left: the thick solid line, the thin solid line, and the dashed line indicate $m^2=-2,  m^2 = 0.31$, and $m^2=1.36$, respectively.
Right: the thick solid line and the thin solid line correspond to the cases $m^2= -3$ and $m^2 = 1.29$. }\label{psi2}
\end{figure}

\begin{figure}
\centering
\includegraphics[width=8cm]{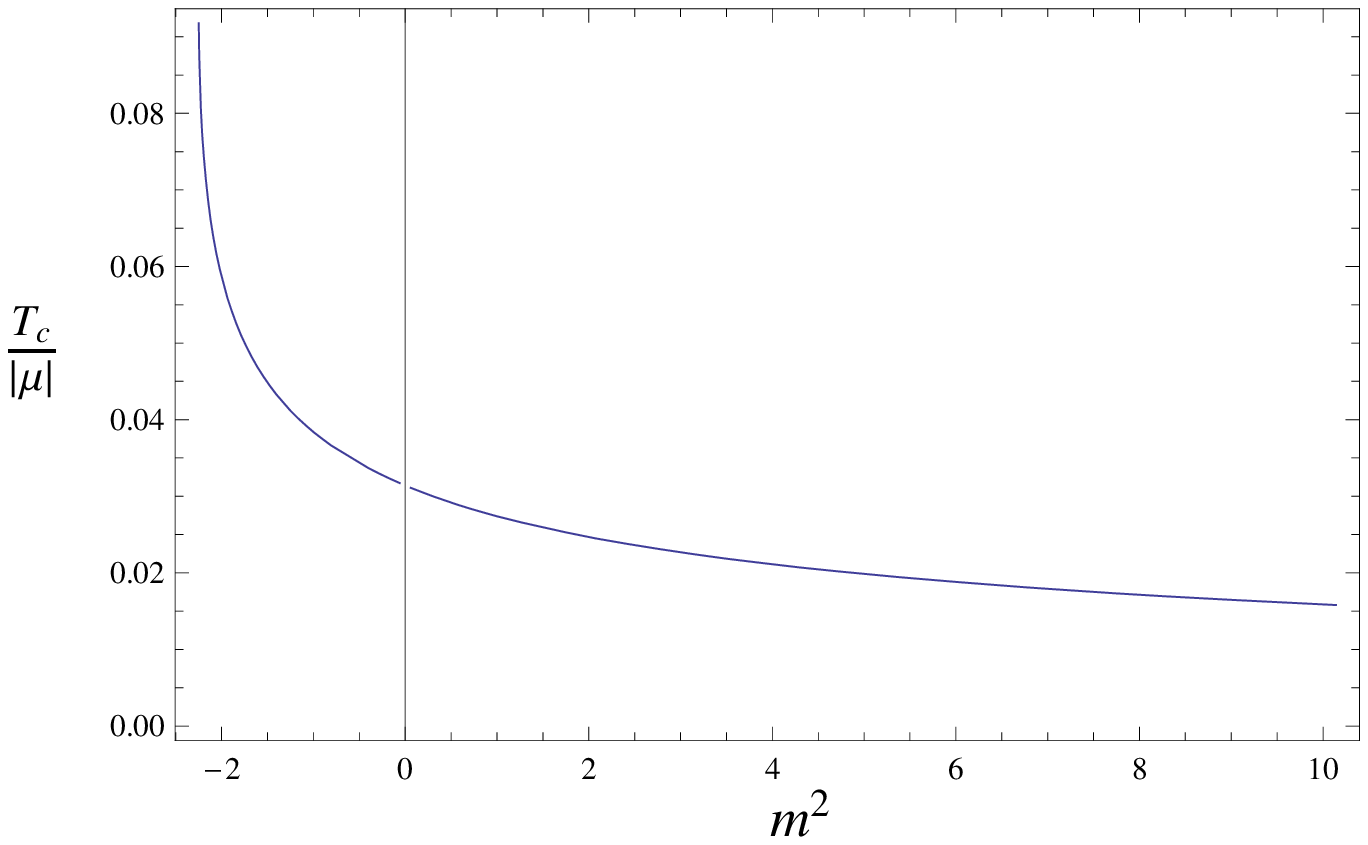}~
\includegraphics[width=8 cm]{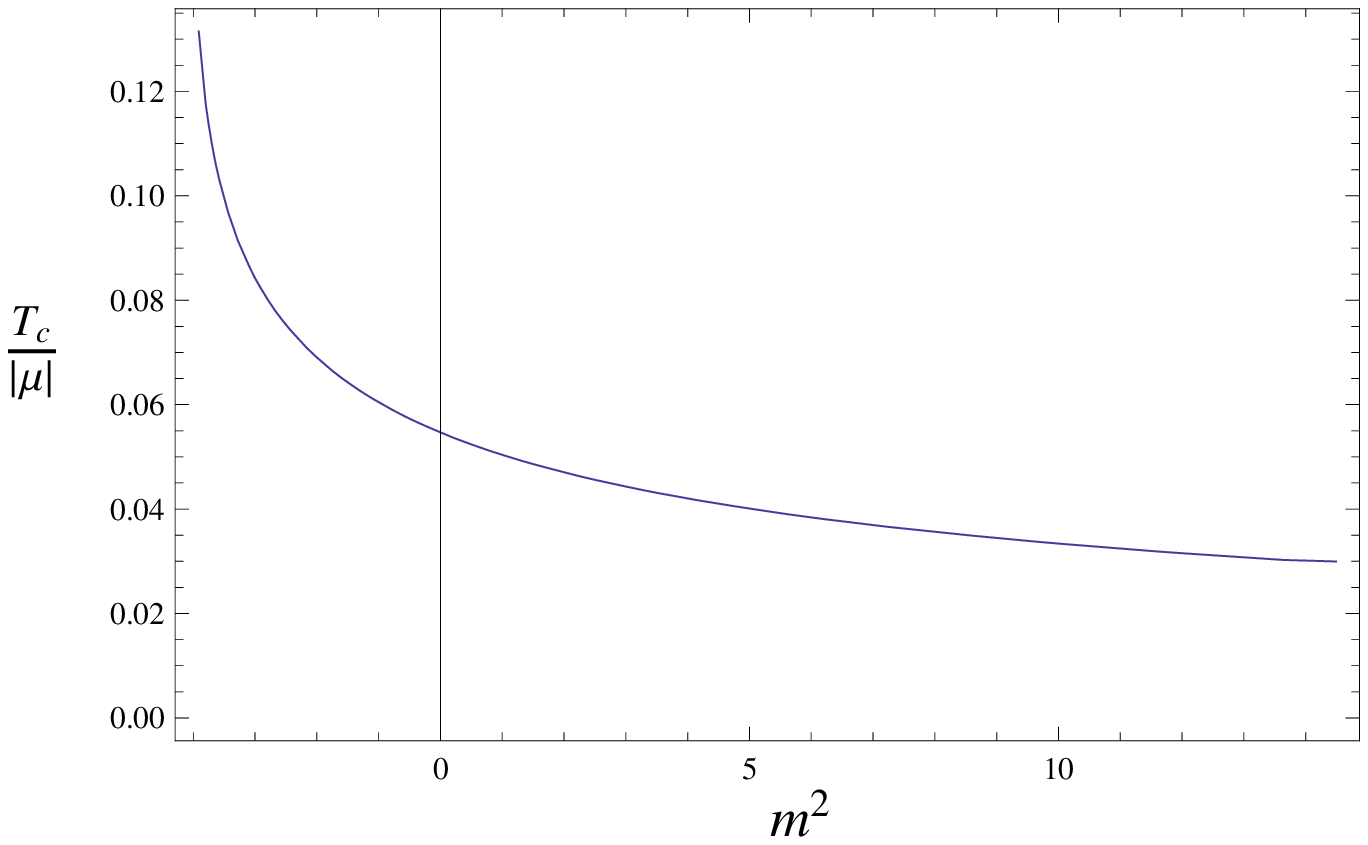}
\caption{\label{tcfig} The critical temperature $T_c$ as a function of $m^2$ in d=2+1 (left) and in d=3+1 (right). }
\end{figure}

Discarding the disordered part, the scalar field condensates for various $m^2$ are shown in Fig. \ref{psi2}. In Fig. \ref{tcfig}, we plot the critical temperature
as a function of the mass squared. $T_c$ decreases as $m^2$ increases, as it should be.


\section{Specific Heat and Conductivity}

In the previous section, we studied the density driven symmetry breaking in particular with large conformal dimensions.
We now consider two observables that exhibit characteristics of superconductors with negative and positive values of $m^2$ to see
how those observables are affected by varying $m^2$ or conformal dimensions.
We first consider the specific heat of holographic superconductors. It is known that the specific heat of superconductors shows an interesting behavior at low and high temperatures: at a temperature far below the transition temperature $T_c$,
the specific heat is exponentially small due to the pairing gap, while it exhibits a discontinuity very near $T_c$.
To calculate the specific heat, let us compute the  free energy of the boundary field theory.

The free energy is obtained by the on-shell action of Eq. (\ref{action})
\begin{equation}
\Omega = - T\, S_{\rm os}+ \cdots\, ,
\end{equation}
where
\begin{eqnarray}
S_{\rm os}= \int d^d x \left[ \frac{\sqrt{-g} g^{zz}}{2} {\left( \frac{1}{e^2} g^{\mu \nu} A_{\mu} A'_{\nu} + \psi \psi' \right)\vline}_{\,\,z=\epsilon}  + \frac{1}{2}\int^{z_0}_{\epsilon} dz \sqrt{-g} A_{\mu} A^{\mu} \psi^2 \right]\, .
\end{eqnarray}
Here $\epsilon$ is introduced to cut off the divergence at the boundary that will be regularized by a boundary counter-term.
With the regularized on-shell action, we obtain the free energy of the superconducting phase
\begin{eqnarray}
\frac{\Omega_s(\mu)}{V} =\frac{1}{2} \left[ (2-d) \mu \rho + \int^{z_0}_{0} \left(\frac{1}{z^{d-1} f} \right) A_t^2 \psi^2 \,dz \right]\, ,
\end{eqnarray}
where $V$ is the volume factor.

\begin{figure}
\centering
\includegraphics[width=7.5 cm]{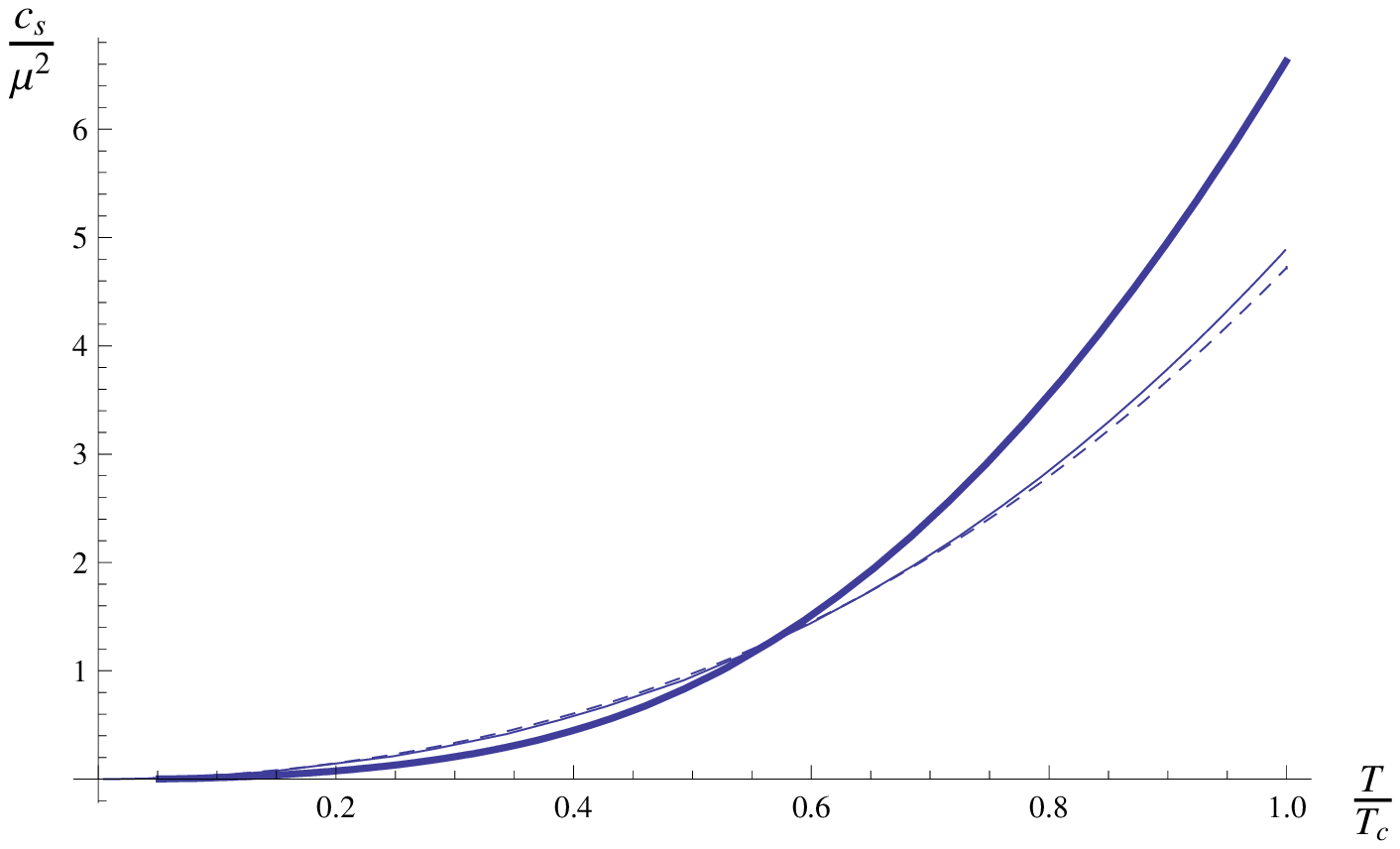}\quad
\includegraphics[width=7.5 cm]{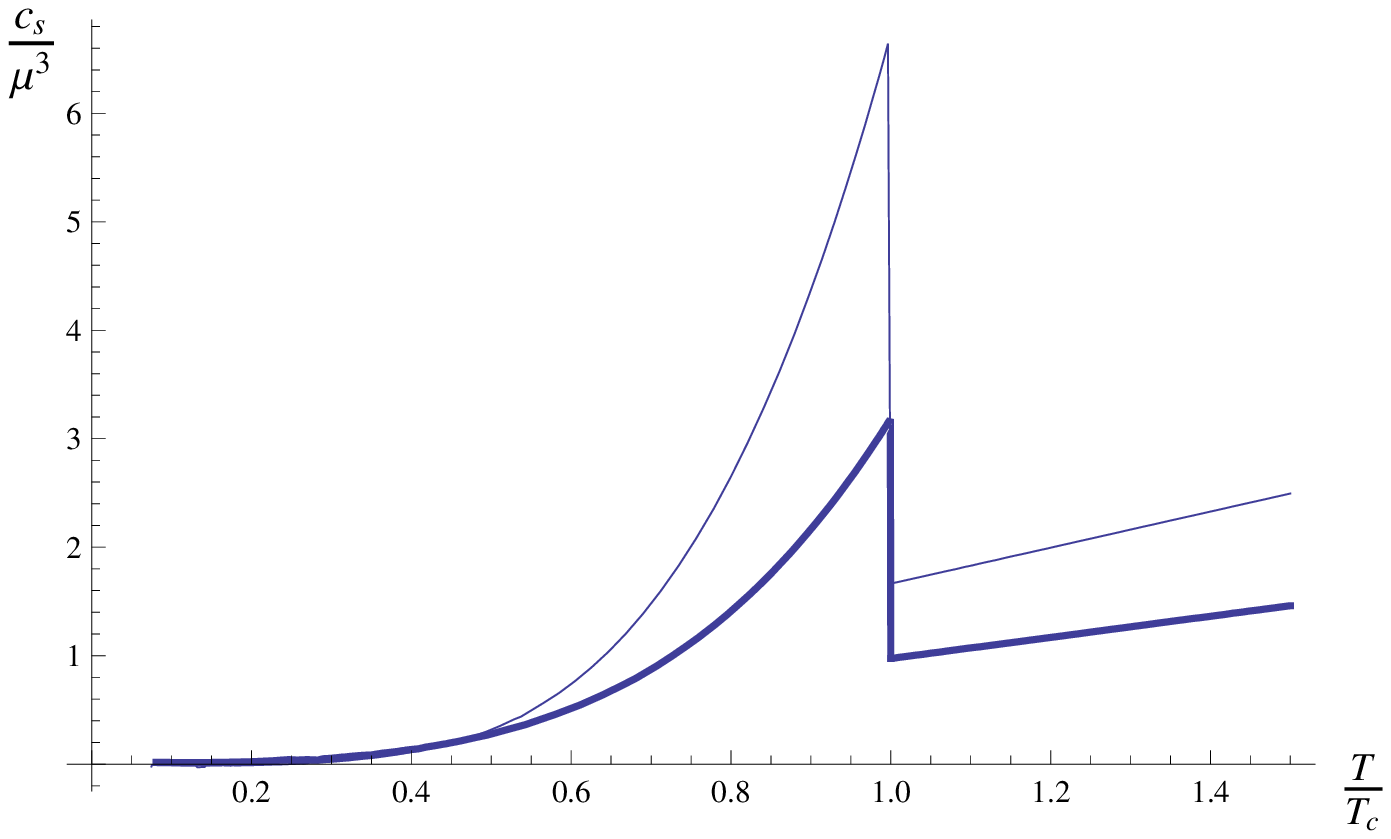}
\caption{\label{sh} Specific heat as a function of temperature in d=2+1(left) and in d=3+1(right). Left : The thick solid line, the thin solid line, the dashed line are for $m^2 = -2, m^2 = 0.31, m^2= 1.36$ respectively. Right : The thin solid line is for $m^2=-3$, and the thick solid line is for $m^2 = 1.29$.}
\end{figure}

{}From the free energy, we obtain the specific heat as
\begin{eqnarray}
c_s= - \frac{T}{V} \frac{\partial^2 \Omega_s}{\partial T^2}\, .
\end{eqnarray}
We note that in experiment, one generally measures the specific heat at constant pressure, $c_p$.
The difference between $c_p$ and $c_v$, the specific heat at fixed volume, is
$\sim (\partial V/\partial T)^2$.  Therefore we may ignore the difference
in condensed matter systems since the volume of the system will  hardly change with respect to the temperature.
In Fig.~\ref{sh}, we plot our results of the specific heat for the holographic superconductors with a few masses.

For usual superconductors, it is known that the specific heat vanishes exponentially as the temperature approaches zero;
\ba
c_s\sim e^{-\frac{\Delta_0}{T}}\, ,
\ea
where $\Delta_0$ is the gap at zero temperature.
In our evaluation of the specific heat, we do not observe the exponential suppression near $T=0$. Due to some numerical instability near zero temperature, it is not clear  whether this is a genuine feature of the holographic superconductors or numerical error.
We notice that a similar observation has made previously in Ref. \cite{Hartnoll:2008kx}.

Now, we consider the ratio of the specific heat of the normal phase ($c_n$)
to the superconducting phase at the critical temperature. Here, normal phase is defined as the one with no condensation. i.e., $\psi(z)=0$.
The ratio is  useful since there are experimentally measured values of it, and also
 near the critical temperature,  we can trust our numerical calculations. 
 The ratio is defined by
\ba
R\equiv {\frac{c_s-c_n}{c_n}\,\,\vline}_{\,\,T=T_c}\, ,
\ea
which measures the discontinuity of the specific heat at the critical temperature.
The calculated ratios are listed in Table 1.
 \begin{center}
 Table 1 : \parbox[t]{5.3in}{The values of $R$ for various $m^2$ for the holographic superconductors in d=3+1. }
  \end{center}
 $$
  \begin{array}{|r|r|}
  \hline
   {\rm m^2}\,\,\, & R\,\, \\
  \hline
  \hline
 {\rm -3} \,\,\,\,\,& 3.01 \,\,\\
  \hline
 {\rm 1.29} \,\,\,\,\,& 2.27 \,\,\\
  \hline
{\rm 1.76} \,\,\,\,\,& 2.24 \,\,\\
  \hline
  {\rm 3.29} \,\,\,\,\,& 2.17 \,\,\\
  \hline
{\rm 5} \,\,\,\,\,& 2.11 \,\,\\
  \hline
\end{array}
  $$

To see if our results have any relevance to a real condensed matter system, we may compare our numbers with the ones from experiments. In Table 2, we list the values of $R$ for some materials \cite{Poole}.

\begin{center}
  Table 2 : \parbox[t]{5.3in}{The values of R for various materials. }
  \end{center}
 $$
  \begin{array}{|r|r||r|r|}
  \hline
   {\rm Material }\,\,\, & R\,\, & {\rm Material}\,\,\,&R\,\, \\
  \hline
  \hline
  {\rm BCS~ theory} \,\,\,\,\,& 1.43 &{\rm Al}\,\,\,\,\,&1.45\\
  \hline
 {\rm Cd} \,\,\,\,\,& 1.36 &{\rm Pb}\,\,\,\,\,&2.71\\
  \hline
  {\rm CeRu_2Si_2} \,\,& 3.5 &{\rm BaPb_{1-x}Bi_x O_3}&8\\
  \hline
  {\rm (La_{0.925}Ca_{0.075})_2CuO_4}& 5.7 &{\rm YBa_2Cu_3O_7}&3.6\\
  \hline
    \end{array}
  $$

\begin{figure}
\centering
\includegraphics[width=8cm]{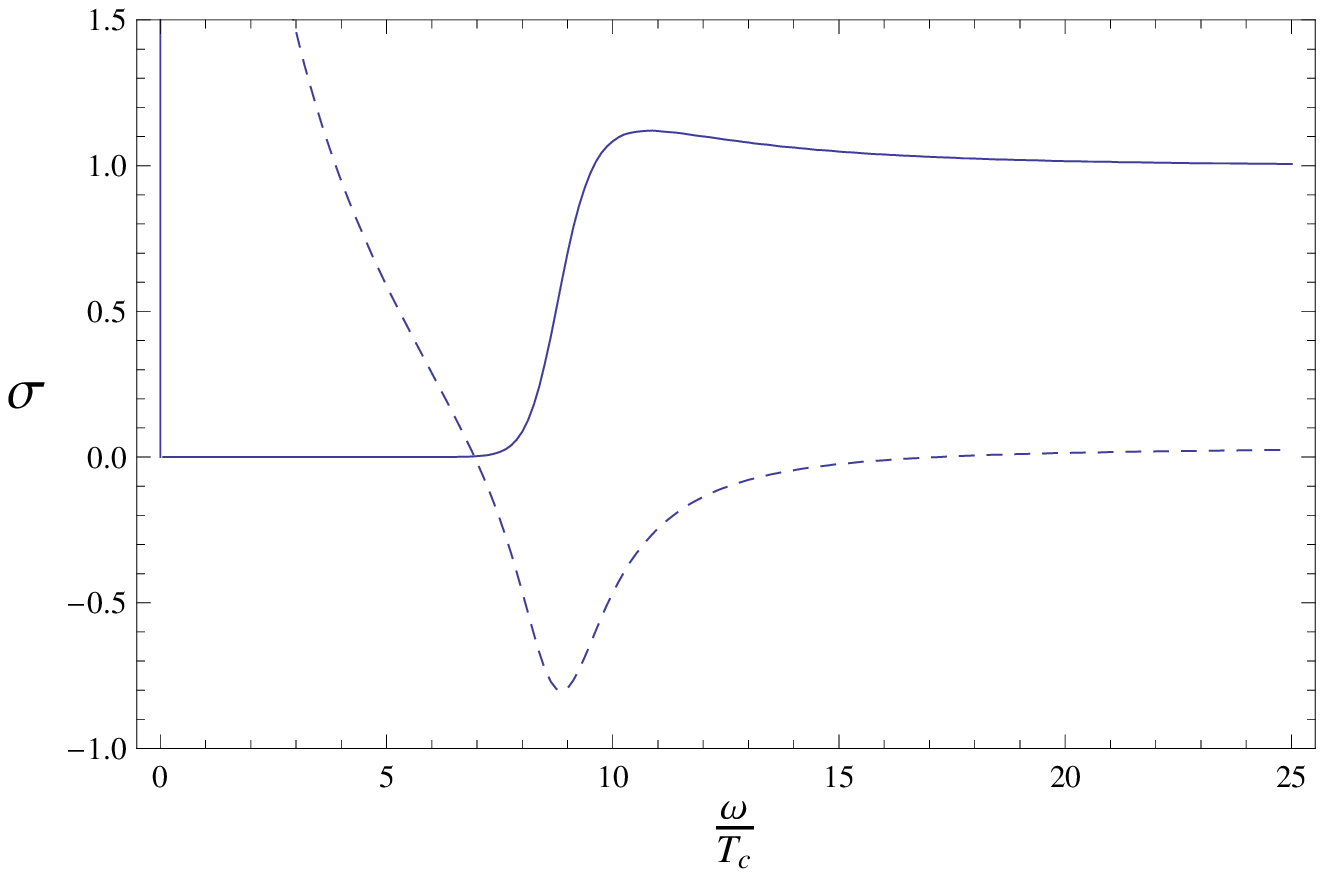}~
\includegraphics[width=8cm]{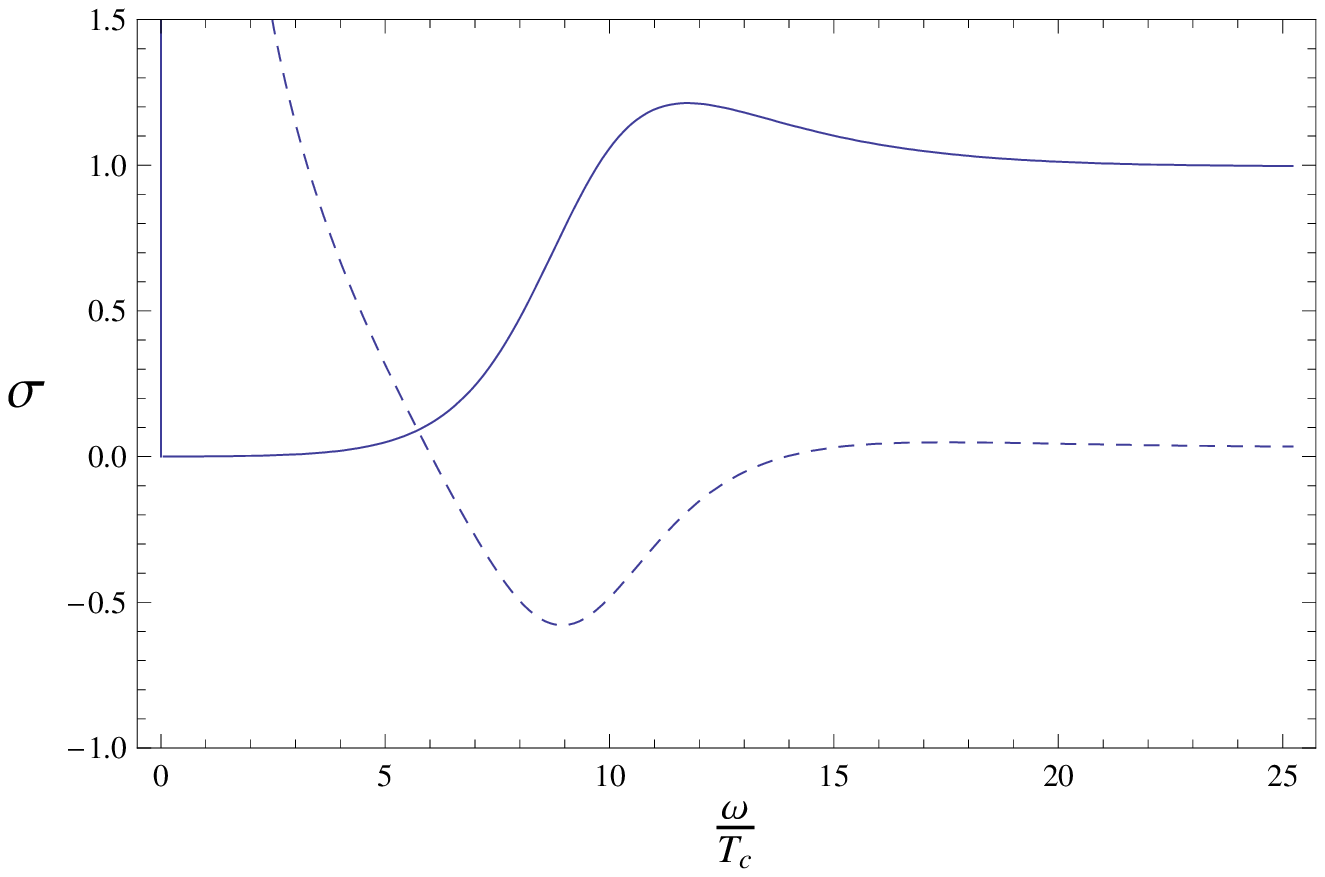}~
\caption{\label{sig2} The conductivity at $T\sim 0.15 \,T_c$ for $m^2=-2$ 
(left), and $m^2= 1.36$ (right).
The solid line is for the real part, while the dashed line represents the imaginary part. }
\end{figure}

Finally, we consider the electrical conductivity, which shows the characteristics of superconductors.
The conductivity can be obtained by taking into account the retarded Green's function,
$\sigma (\omega)= \frac{1}{i\omega}G^{\rm R}(\omega, k=0)$.
Now we introduce a gauge field $A_x$ in the bulk as a fluctuation which couples to an external electric current at the boundary.
The equation of motion for $A_x$ is
\begin{equation}
A_x'' + \left( \frac{f'}{f} + \frac{3-d}{z}\right) A_x' +\left( \frac{\omega^2}{f^2} - \frac{2 \,\psi^2}{f z^2 }\right) A_x = 0\, .
\end{equation}
To solve the equation, we impose the ingoing boundary condition, $A_x \sim f^{- i \omega z_0/d}$ at the horizon.
At the boundary, we have
\begin{equation}
A_x = A^{(0) } + A^{(1)} z + \cdots\, .
\end{equation}
Then the retarded Green's function reads
\ba
G^R=\frac{A^{(1)}}{A^{(0)}}\, ,
\ea
and thereby
\begin{eqnarray}
\sigma (\omega) = -\frac{i\, A_x^{(1)}}{\,\omega\, A_x^{(0)}}\mid_{k=0}\, .
\end{eqnarray}
In Fig.~\ref{sig2}, we present  conductivities for $m^2=-2,$ and $m^2 = 1.36$ in 2+1 dimensions. We do not find any qualitative difference between the positive and negative $m^2$ cases.
Our results are qualitatively similar to the one obtained previously with $m^2\le 0$~\cite{Horowitz:2008bn, Hartnoll:2008kx}.
The peak of the conductivity becomes a little bit broad with increasing $m^2$, but the dependence is weak.

\section{Summary and Discussion}
 We studied the holographic superconductors by the density driven instability by considering  bulk scalar fields with $m^2>0$. We found that the holographic superconductors still exist with  positive $m^2$. We observed that as the bulk mass of the scalar field  increases, the system exhibits the phase space folding: two very close boundary conditions  can give completely opposite states; one  symmetry  broken state, and  the other  symmetric state.
Such sharp dependence on  boundary conditions is reminiscent of ``the butterfly effect'', a road to chaos.
As a consequence, for large conformal operators, the initial conditions prepared at the horizon should be highly fine tuned to obtain the density driven symmetry breaking. However, we remark here that while the usual butterfly effect is in  the time direction, our case is for a radial evolution in the  bulk of the dual gravity. It would be interesting to study it further in relation to the chaos \cite{chaos}.

We also studied other quantitative properties of the holographic superconductors, i.e. the specific heat and the conductivity.
We found that the value of the specific heat increases rapidly as temperature approaches the critical temperature, which agrees with a genuine feature of superconductivity in real materials.
The ratio of the specific heat of the normal state to the superconducting state at the critical temperature are
 compared  with experimentally observed ones to see if holographic superconductors share some common feature with real materials.
 The obtained behavior of the conductivity was rather universal against varying $m^2$.

Finally, we remark that to establish the instability of the
folded region in the phase space more reliably, one needs to study the back reaction of the scalar field to the geometry. The instability for the large chemical potential sectors might be due to the fixed metric.
 In this work, the charge is allowed to fluctuate. If the scalar field were absent, we know that the charge is not supposed to be larger than  the critical value  when considering a fixed mass. Here, the coupling of the gauge field and the scalar field makes things fuzzy.
The work \cite{Denef} claims that for any values of the scalar charge larger than a bound value, the condensate is always possible. It would be interesting to study how our work  is related  to it, and the work in this direction is in preparation.

\begin{acknowledgments}
We want to thank P. Basu, S. Hartnoll and A. Yarom for useful comments
on the first version of the paper.
This work is supported in part
by the National Research Foundation of Korea(NRF) grant funded by the Korea government(MEST) (No. 20090063068).
The work of S.J. Sin was also supported in part by KOSEF Grant R01-2007-000-10214-0. 
 Y. Kim acknowledges the Max Planck Society(MPG), the Korea Ministry of Education, Science, Technology(MEST), Gyeongsangbuk-Do and Pohang City for the support of the Independent Junior Research Group at the Asia Pacific Center for Theoretical Physics(APCTP).

\end{acknowledgments}


\end{document}